\documentclass{aastex}
\usepackage{amsmath}
\usepackage{amssymb}
\usepackage{textcomp}
\usepackage{epstopdf}
\shorttitle{Trapped protons measured by PAMELA}
\shortauthors{Adriani et al.}

\begin{document}

\title{Trapped proton fluxes at low Earth orbits\\measured by the PAMELA experiment}

\author{
O.~Adriani$^{1,2}$,
G.~C.~Barbarino$^{3,4}$,
G.~A.~Bazilevskaya$^{5}$,
R.~Bellotti$^{6,7}$,
M.~Boezio$^{8}$,
E.~A.~Bogomolov$^{9}$,
M.~Bongi$^{1,2}$,
V.~Bonvicini$^{8}$,
S.~Bottai$^{2}$,
A.~Bruno$^{6,*}$,
F.~Cafagna$^{7}$,
D.~Campana$^{4}$,
R.~Carbone$^{8}$,
P.~Carlson$^{10}$,
M.~Casolino$^{11,12}$,
G.~Castellini$^{13}$,
C.~De Donato$^{11,15}$,
C.~De Santis$^{11,15}$,
N.~De Simone$^{11}$,
V.~Di Felice$^{11,16}$,
V.~Formato$^{8,17}$,
A.~M.~Galper$^{14}$,
A.~V.~Karelin$^{14}$,
S.~V.~Koldashov$^{14}$,
S.~Koldobskiy$^{14}$,
S.~Y.~Krutkov$^{9}$,
A.~N.~Kvashnin$^{5}$,
A.~Leonov$^{14}$,
V.~Malakhov$^{14}$,
L.~Marcelli$^{11,15}$,
M.~Martucci$^{15,18}$,
A.~G.~Mayorov$^{14}$,
W.~Menn$^{19}$,
M.~Merg$\acute{e}$$^{11,15}$,
V.~V.~Mikhailov$^{14}$,
E.~Mocchiutti$^{8}$,
A.~Monaco$^{6,7}$,
N.~Mori$^{1,2}$,
R.~Munini$^{8,17}$,
G.~Osteria$^{4}$,
F.~Palma$^{11,15}$,
B.~Panico$^{4}$,
P.~Papini$^{2}$,
M.~Pearce$^{10}$,
P.~Picozza$^{11,15}$,
M.~Ricci$^{18}$,
S.~B.~Ricciarini$^{2,13}$,
R.~Sarkar$^{20,21}$,
V.~Scotti$^{3,4}$,
M.~Simon$^{19}$,
R.~Sparvoli$^{11,15}$,
P.~Spillantini$^{1,2}$,
Y.~I.~Stozhkov$^{5}$,
A.~Vacchi$^{8}$,
E.~Vannuccini$^{2}$,
G.~I.~Vasilyev$^{9}$,
S.~A.~Voronov$^{14}$,
Y.~T.~Yurkin$^{14}$,
G.~Zampa$^{8}$,
N.~Zampa$^{8}$,
and V.~G.~Zverev$^{14}$
}

\affil{$^{1}$ Department of Physics and Astronomy, University of Florence, I-50019 Sesto Fiorentino, Florence, Italy}
\affil{$^{2}$ INFN, Sezione di Florence, I-50019 Sesto Fiorentino, Florence, Italy}
\affil{$^{3}$ Department of Physics, University of Naples ``Federico II'', I-80126 Naples, Italy}
\affil{$^{4}$ INFN, Sezione di Naples, I-80126 Naples, Italy}
\affil{$^{5}$ Lebedev Physical Institute, RU-119991 Moscow, Russia}
\affil{$^{6}$ Department of Physics, University of Bari, I-70126 Bari, Italy}
\affil{$^{7}$ INFN, Sezione di Bari, I-70126 Bari, Italy}
\affil{$^{8}$ INFN, Sezione di Trieste, I-34149 Trieste, Italy}
\affil{$^{9}$ Ioffe Physical Technical Institute, RU-194021 St. Petersburg, Russia}
\affil{$^{10}$ KTH, Department of Physics, and the Oskar Klein Centre for Cosmoparticle Physics, AlbaNova University Centre, SE-10691 Stockholm, Sweden}
\affil{$^{11}$ INFN, Sezione di Rome ``Tor Vergata'', I-00133 Rome, Italy}
\affil{$^{12}$ RIKEN, Advanced Science Institute, Wako-shi, Saitama, Japan}
\affil{$^{13}$ IFAC, I-50019 Sesto Fiorentino, Florence, Italy}
\affil{$^{14}$ National Research Nuclear University MEPhI, RU-115409 Moscow, Russia}
\affil{$^{15}$ Department of Physics, University of Rome ``Tor Vergata'', I-00133 Rome, Italy}
\affil{$^{16}$ Agenzia Spaziale Italiana (ASI) Science Data Center, Via del Politecnico snc, I-00133 Rome, Italy}
\affil{$^{17}$ Department of Physics, University of Trieste, I-34147 Trieste, Italy}
\affil{$^{18}$ INFN, Laboratori Nazionali di Frascati, Via Enrico Fermi 40, I-00044 Frascati, Italy}
\affil{$^{19}$ Department of Physics, Universit\"at Siegen, D-57068 Siegen, Germany}
\affil{$^{20}$ Indian Centre for Space Physics, 43 Chalantika, Garia Station Road, Kolkata 700084, West Bengal, India}
\affil{$^{21}$ Previously at INFN, Sezione di Trieste, I-34149 Trieste, Italy. }

\altaffiltext{*}{Corresponding author. E-mail address: alessandro.bruno@ba.infn.it.}

\begin{abstract}
We report an accurate measurement of the geomagnetically trapped proton fluxes for kinetic energy above $\sim$ 70 MeV
performed by the PAMELA mission at low Earth orbits (350 $\div$ 610 km). Data were analyzed in the frame of the
adiabatic theory of charged particle motion in the geomagnetic field. Flux properties were investigated in detail,
providing a full characterization of the particle radiation in the South Atlantic Anomaly region, including locations,
energy spectra, and pitch angle distributions. PAMELA results significantly improve the description of the Earth's
radiation environment at low altitudes, placing important constraints on the trapping and interaction processes, and
can be used to validate current trapped particle radiation models.

\end{abstract}

\keywords{astroparticle physics --- atmospheric effects --- cosmic rays --- elementary particles --- magnetic fields --- space vehicles}

\section{Introduction}\label{Introduction}
The radiation or Van Allen belts are regions of the Earth's magnetosphere where energetic charged particles experience long-term magnetic trapping.
The outer belt is predominately populated by electrons with hundreds of keV to MeV energies. The inner belt consists of an intense radiation of energetic protons (up to a few GeV), with a minor component of $e^{\pm}$ and ions. Protons with energies greater than some tens of MeV mainly originate from the $\beta$-decay of free neutrons produced in the interaction of galactic cosmic rays (CRs) with the Earth's atmosphere, according to the so-called ``Cosmic Ray Albedo Neutron Decay'' (CRAND) mechanism \citep{Singer,Farley}.

The most widespread empirical trapped proton model in last few decades is the NASA AP8 model \citep{AP8}, a static global map of long-term average trapped proton flux, based on a series of measurements performed in the 1960s and early 1970s;
two versions were developed for maximum and minimum solar conditions, respectively.
Similar models were provided by the Institute of Nuclear Physics of Moscow State University (INP/MSU, \citealt{INP}).
Recently, significant improvements \citep{CRRESPRO,Huston,PSB97,TPM1} have been made due to the data from new satellite experiments, such as CRRES \citep{CRRES,CRRES2}, SAMPEX/PET \citep{Looper96,Looper98} and the TIROS/NOAA series \citep{Huston96}. Nevertheless,
the modeling of the low-altitude radiation environment is still incomplete, with
largest uncertainties affecting the high-energy ($>$ 50 MeV) fluxes in the inner zone and the South Atlantic Anomaly (SAA), where the inner belt makes its closest approach to the Earth's surface\footnote{The SAA is a consequence of the tilt ($\sim$10 deg) between the magnetic dipole axis of the Earth and its rotational axis, and of the offset ($\sim$500 km) between the dipole and the Earth centers. This region is characterized by the extremely low intensity of the geomagnetic field, with a significant contribution from non-dipolar components.}.

New accurate measurements of the high-energy ($\gtrsim$ 70 MeV) CR radiation in Low Earth Orbits (LEO) have been reported by the PAMELA mission \citep{Picozza}.
Due to the orbit and the high identification capabilities, PAMELA is able to provide detailed information about particle fluxes in different regions of the terrestrial magnetosphere, including energy spectra and spatial and angular distributions.
In particular, the spacecraft passes through the SAA, allowing the observation of geomagnetically trapped particles from the inner Van Allen belt. For the first time PAMELA has revealed the existence of a significant component of trapped antiprotons in the inner
belt \citep{PAMELAtrappedpbars}. In this Letter, we present the measurement of the trapped proton fluxes.

\section{PAMELA Data Analysis}\label{PAMELA data analysis}

PAMELA is a space-based experiment designed for a precise measurement of the charged cosmic radiation in the kinetic energy range from some tens of MeV up to several hundreds of GeV.
The \emph{Resurs-DK1} satellite, which hosts the apparatus, has a semi-polar (70 deg inclination) and elliptical (350$\div$610 km altitude) orbit. 
The spacecraft is three-axis stabilized. The orientation is calculated by an on board processor with an accuracy better than 1 deg which, together with the good angular resolution ($<$ 2 deg) of PAMELA, allows particle direction to be measured with high precision.

The analyzed data set includes protons acquired by PAMELA between 2006 July and 2009 September.
In order to account for the time variations of PAMELA detector performance, with the major effect related to the sudden failure of
some front-end chips in the tracking system, data were divided into sub-sets;
consistent with the spacecraft orbit precession rate, the sub-set width was chosen to be of about 244 days.
Measured rigidities were corrected for the energy loss in the apparatus with MonteCarlo simulations.
Details about apparatus performance, particle selection, efficiencies and measurement uncertainties can be found elsewhere \citep{ProHe,SOLARMOD}.

Data were analyzed in the framework of the adiabatic theory, which provides a relatively simple description of the complex dynamics of charged particles in the magnetosphere. The motion of trapped particles was assumed to be a superposition of three periodic motions: a gyration around the local magnetic field lines, a bouncing along field lines between conjugate mirror points in the northern and southern magnetic hemispheres, and a drift around the Earth. Each type of motion is related to an adiabatic invariant, which is conserved under the condition of small magnetic field variations during the period of the motion, and in absence of energy loss, nuclear scattering and radial diffusion \citep{Walt}.
In order to reject events near the local geomagnetic cutoff, characterized by chaotic trajectories of non-adiabatic type, only protons with rigidities $R<10/L^{3}$ GV were selected, where $L$ is the McIlwain's parameter \citep{McIlwain} measured in Earth's radii ($R_{E}$=6371.2 km).

McIlwain's coordinates and other variables of interest, as the adiabatic invariants, were calculated on an event-by-event basis using
the IRBEM library \citep{IRBEM}.
The IGRF-10 \citep{IGRF10} and the TS05 \citep{TS05} models were used for the description of the internal and external geomagnetic field, respectively: the former employs a global spherical harmonic implementation of the main magnetic field; the latter is a dynamical (five-minute resolution) model of the storm-time geomagnetic field in the inner magnetosphere, based on recent satellite measurements.

\subsection{Trajectory Reconstruction}
Proton trajectories were reconstructed in the Earth's magnetosphere using a tracing program based on numerical integration methods \citep{TJPROG,SMART}, and implementing the aforementioned geomagnetic field models. For each event, the number of gyrations, bounces and drifts was evaluated in order to estimate corresponding frequencies and check trajectory behaviors. Trajectories were propagated back and forth from the measurement location, and traced until:
\begin{enumerate}
  \item they reached
  the model magnetosphere boundaries;
  \item or they intersected the absorbing atmosphere limit, which was assumed at an altitude\footnote{Such a value approximately corresponds to the mean production altitude for protons created in CR interactions on the atmosphere.} of 40 km;
  \item or they performed more than $10^{6}/R^{2}$ steps, where $R$ is the particle rigidity in GV, for both propagation directions.
\end{enumerate}
Case (1) corresponds to protons from interplanetary space possibly surviving the cutoff selection $R<10/L^{3}$, which were excluded from the analysis.
Events satisfying condition (3) were classified as stably-trapped protons:
since the program uses a dynamic variable step length, which is of the order of 1\% of a particle gyro-distance in the magnetic field,
the applied rigidity-dependent criterion ensures that at least four drift cycles around the Earth were performed.
Their trajectories were verified to fulfill the adiabatic conditions, in particular the hierarchy of temporal scales:
\begin{equation}\label{hierarchy}
\omega_{gyro} \gg \omega_{bounce} \gg \omega_{drift},
\end{equation}
where $\omega_{gyro}$, $\omega_{bounce}$ and $\omega_{drift}$ are the frequencies associated to the gyration, the bouncing and the drift motion, respectively;
estimated frequency ratios $\omega_{bounce}/\omega_{gyro}$ and $\omega_{drift}/\omega_{bounce}$ are of the order of $10^{-2}\div10^{-1}$.
Finally, category (2) includes quasi-trapped and un-trapped protons with limited lifetimes: the former have trajectories similar to those of stably-trapped protons, but originated and are re-absorbed by the atmosphere
during a time larger than a bounce period (up to several tens of seconds); conversely, the latter precipitate into the atmosphere within a bounce period ($\lesssim$ 1 s). The analysis of such components
is beyond the aim of this work and it will be
the subject of a specific paper.

The trajectory analysis
allows
a deeper
investigation
of proton populations in LEO, including energy-dependent effects
such as
the breakdown of the trapping mechanism at high energies as a consequence of either large gyro-radius or non-adiabatic trajectory effects \citep{Selesnick2007}.

\subsection{Flux Calculation}
The factor of proportionality between fluxes and numbers of detected particles, corrected for selection efficiencies and acquisition time, is by definition the apparatus gathering power $\Gamma$.
For an isotropic particle flux, the gathering power depends only on the detector design, and it is usually called the geometrical factor $G_{F}$. In the case of the PAMELA apparatus $\Gamma$ (and $G_{F}$) also depends on particle rigidity,
due to the spectrometer bending effect on particle trajectories.
Conversely, in the presence of anisotropic fluxes, the gathering power depends on the direction of observation as well. Consequently, an accurate estimate of $\Gamma$ is crucial in evaluating fluxes inside and near the SAA region, where the trapped radiation is highly anisotropic as a consequence of the interaction with the Earth's atmosphere.

Similar to \citet{Selesnick}, the
PAMELA effective area (cm$^{2}$) was evaluated as a function of particle energy $E_{k}$, local pitch angle $\alpha$ (i.e the angle between particle velocity vector and geomagnetic field line)
and satellite orientation $\Psi$ with respect to the geomagnetic field:
\begin{equation}\label{area_formula}
H(E_{k},\alpha,\Psi)=\frac{1}{2\pi}\int_{0}^{2\pi} d\beta \left[ A(E_{k},\theta,\phi) \cdot sin\alpha \cdot cos\theta \right],
\end{equation}
where $\beta$ is the gyro-phase angle and $\theta$=$\theta(\alpha,\beta,\Psi)$ and $\phi$=$\phi(\alpha,\beta,\Psi)$ are, respectively, the zenith and the azimuth angle describing particle direction in the PAMELA frame\footnote{The PAMELA reference system originated in the center of the spectrometer cavity; the Z axis is directed along the main axis of the apparatus toward the incoming particles; the Y axis is directed opposite to the main direction of the magnetic field inside the spectrometer; the X axis completes a right-handed system.}, and $A(E_{k},\theta,\phi)$ is the apparatus response function \citep{Sullivan}.
For isotropic fluxes
$H$ does not depend on $\Psi$ and it is related to the geometrical factor $G_{F}(E_{k})$ by:
\begin{equation}\label{gf_formula}
G_{F}(E_{k})=2\pi  \int_{0}^{\pi}d\alpha H(E_{k},\alpha).
\end{equation}

The effective area of the PAMELA apparatus
was evaluated with Monte Carlo integration methods \citep{Sullivan},
averaging it over $\beta$ angle.
The calculation was performed by varying $\alpha$ in steps of 1 deg in the range 0$\div$180 deg. The satellite orientation $\Psi=(\theta_{\Psi},\phi_{\Psi})$, where $\theta_{\Psi}$ and $\phi_{\Psi}$ denote, respectively, the zenith and the azimuth angles\footnote{Note that the PAMELA apparatus is not cylindrically symmetric.} of the geomagnetic field direction in the PAMELA reference frame, was varied in steps of $\Delta\theta_{\Psi}$, $\Delta\phi_{\Psi}$=1 deg
over the $\Psi$ domain covered by the spacecraft.
The dependency of the instrument response on proton energy was studied by estimating the effective area in 40 logarithmic bins
($E_{k}$ = 63 MeV $\div$ 40 GeV).

In order to reduce statistical fluctuations due to the limited counts available in each $(E_{k},\alpha,\Psi)$ bin, a mean effective area was derived
at each spacecraft geographic position $\textbf{X}$ = $(Lat, Lon, Alt)$:
\begin{equation}\label{mean_area_formula}
H(\textbf{X},E_{k},\alpha) = \frac{\sum\limits_{\Psi\rightarrow \textbf{X}}  H(E_{k},\alpha,\Psi) \cdot T(\Psi) }{\sum\limits_{\Psi\rightarrow \textbf{X}}  T(\Psi)},
\end{equation}
by weighting each area contribution by the livetime spent by PAMELA at satellite orientations corresponding to $\textbf{X}$.

Differential directional fluxes ($GeV^{-1}m^{-2}sr^{-1}s^{-1}$) were calculated over a five-dimensional grid $F(\textbf{X},E_{k},\alpha)$:
\begin{equation}\label{flux_formula}
F(\textbf{X},E_{k},\alpha)=\frac{N(\textbf{X},E_{k},\alpha)}{2\pi \cdot  H(\textbf{X},E_{k},\alpha) \cdot T(\textbf{X})  \cdot \Delta\alpha \cdot \Delta E_{k}},
\end{equation}
where $N(\textbf{X},E_{k},\alpha)$ is the number of counts corrected for selection efficiencies,
and $T(\textbf{X})=\sum\limits_{\Psi\rightarrow \textbf{X}}  T(\Psi)$ is the total livetime spent at $\textbf{X}$.
The flux grid
extends over the whole phase-space region $(\textbf{X},E_{k},\alpha)$ covered by PAMELA, with $\textbf{X}$ resolution given by $\Delta Lat, \Delta Lon$ = 2 deg and $\Delta Alt$ = 20 km, for a total number of bins amounting to
$\sim10^{8}$.

\subsection{East-West Effect Correction}
Above several tens of MeV finite gyro-radius effects become not negligible.
In particular, at PAMELA energies the proton gyro-radius
can be large up to several hundreds of kilometers,
so that fluxes measured at a given position correspond to particles with different guiding centers (i.e., the center of gyration) locations $\textbf{X}_{gc}$ = $(Lat_{gc}, Lon_{gc}, Alt_{gc})$.
The atmospheric density averaged over a circle of gyration can be appreciably different from the density at the guiding center.
The flux of protons arriving from the East is lower than the flux of protons from the West, since their guiding centers are located at lower altitudes and thus their flux is reduced by the atmospheric absorption. Such a phenomenon, known as the East-West effect \citep{EASTWEST}, was taken into account
by evaluating
fluxes at
corresponding guiding center coordinates.
However, the resulting correction is relatively small because of the limited apparatus aperture and since the PAMELA major axis is mostly oriented toward the zenith direction.

\subsection{Flux Mapping}
At a later stage,
the geographic flux grid $F(\textbf{X}_{gc},\alpha,E_{k})$ was interpolated onto magnetic coordinates, using several invariant coordinate systems. In particular, distributions were evaluated as a function of
adiabatic invariants, providing a convenient description
of trapped fluxes. The versions of the adiabatic invariants used are:
\begin{equation}\label{1inv}
M=\frac{ p^{2} }{ 2 m_{0} B_{m}},
\end{equation}
\begin{equation}\label{2inv}
K=\int_{s_{m}}^{s'_{m}} [B_{m}-B(s)]^{1/2}ds,
\end{equation}
\begin{equation}\label{3inv}
\Phi = \oint\textbf{A} \cdot \textbf{dl},
\end{equation}
where $p$ is momentum, $m_{0}$ is the proton rest mass, $B$ is the
local magnetic field, $B_{m}$ is the mirror point magnetic field,
$s$ is distance along the magnetic field line, the integration is
along the magnetic field between the mirror point locations $s_{m}$
and $s'_{m}$, $\textbf{A}$ is the magnetic vector potential, and the integration is along a curve that lies in the particle drift shell \citep{Roederer}.
$M$, $K$, and $\Phi$ are related to the particle gyration, bounce and drift motion, respectively.
It should be noted that the adiabatic invariants cannot be treated as spatial coordinates since they are properties of the particles.

Alternatively, the particle energy $E_{k}$ was used in place of $M$, and the Roederer parameter was used as a drift invariant:
\begin{equation}\label{3inv2}
L^{*} = \frac{2\pi \mu_{E}}{R_{E}\Phi},
\end{equation}
where $\mu_{E}$ is the Earth's magnetic dipole moment \citep{Roederer}. Note that in contrast to $\Phi$, a constant dipole moment is necessary for $L^{*}$ to be invariant
due to secular geomagnetic variations:
accordingly, PAMELA fluxes were calculated for the 1 January 2008.

Finally, in order to investigate the particle anisotropy, fluxes were also mapped as a function of equatorial pitch angle $\alpha_{eq}$ and McIlwain's $L$-shell.
Such coordinates
are commonly used in the literature as bounce and drift invariants since they provide a more intuitive mapping, so they are useful for comparison with other data sets.

\begin{figure}[!t]
\centering
\includegraphics[width=6.5in]{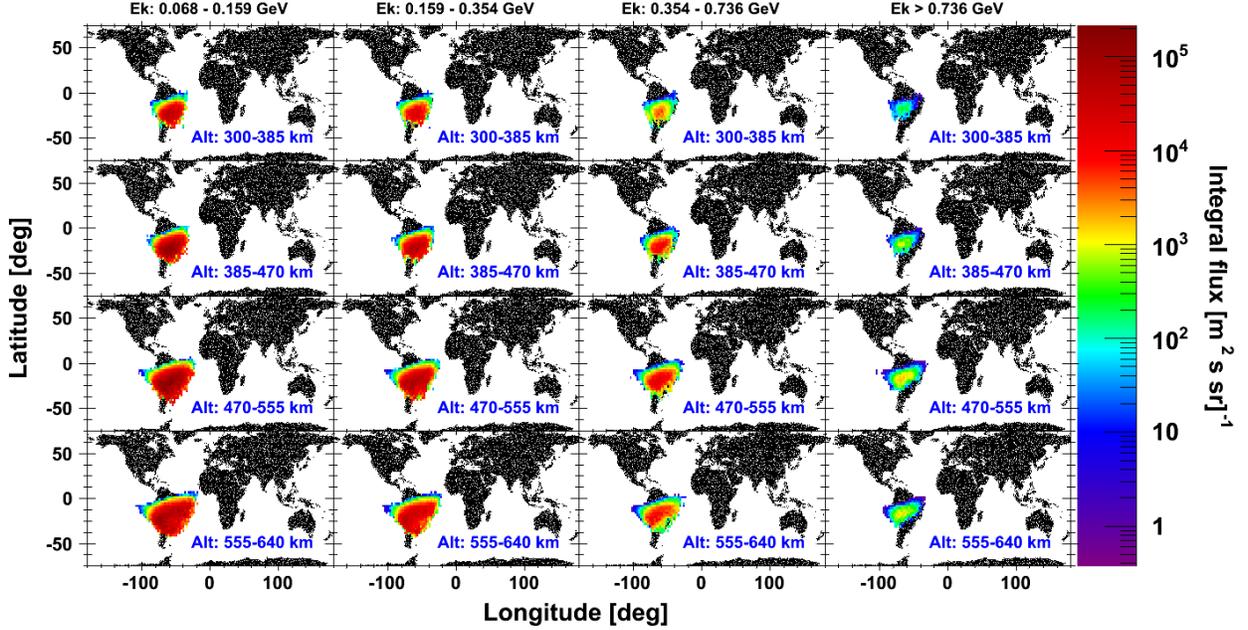}
\caption{Stably-trapped integral fluxes ($m^{-2} s^{-1} sr^{-1}$) averaged over the pitch angle range covered by PAMELA, as a function of geographic coordinates, evaluated for different energy (columns) and guiding center altitude (rows) bins. }
\label{Fig1}
\end{figure}

\section{Results}
The selected sample amounts to $\sim9\cdot10^{6}$ events, including $\sim7.3\cdot10^{6}$ stably-trapped, $\sim5.4\cdot10^{5}$ quasi-trapped and $\sim1.2\cdot10^{6}$ un-trapped protons.

Geographic maps
of the stably-trapped component are shown in Figure \ref{Fig1}, for different bins of kinetic energy (columns) and guiding center altitude (rows). Fluxes were averaged over the local pitch angle range
covered by PAMELA.
Protons from the inner belt are detectable only in the SAA at such altitudes, spreading over a region located over South America.
In particular, they concentrate in the Southeast part of the SAA. The extension of the covered area rapidly changes with altitude and energy.
PAMELA is able to measure trapped fluxes up to their highest energies
($\sim$ 4 GeV).

\begin{figure}[!t]
\centering
\includegraphics[width=6.5in]{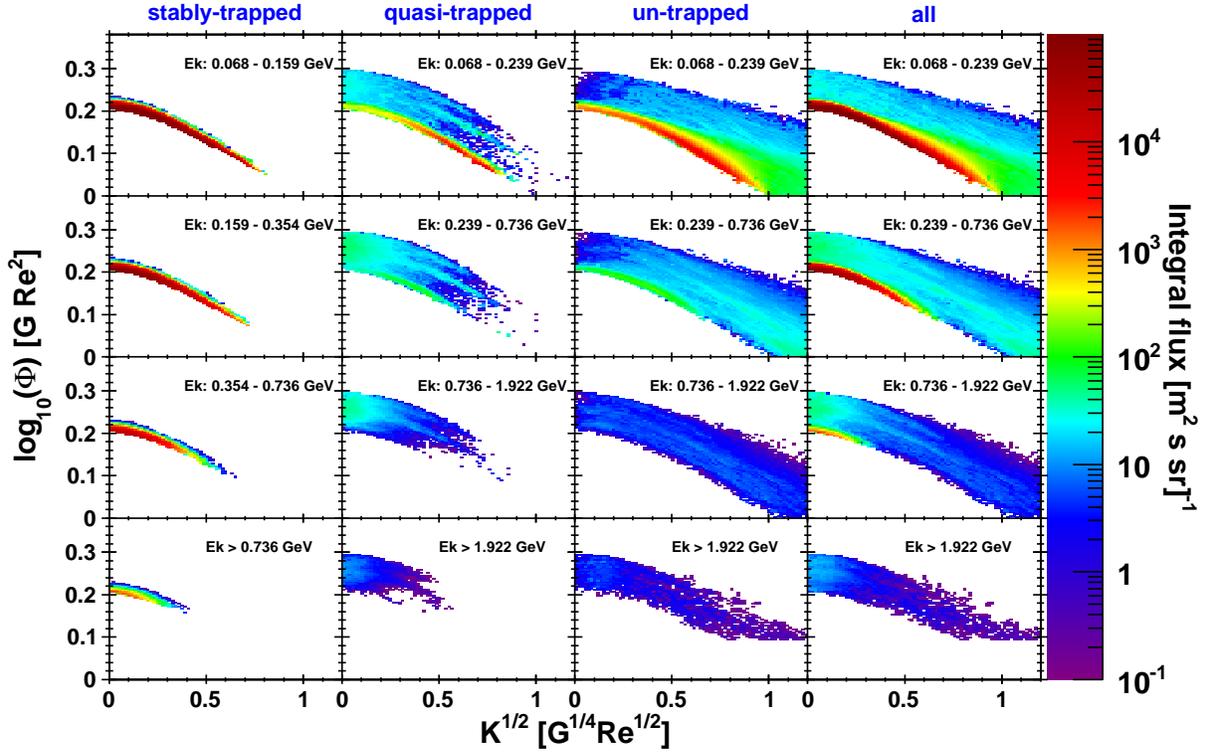}
\caption{Proton integral fluxes ($m^{-2} s^{-1} sr^{-1}$) as a function of the second $K$ and the third $\Phi$ adiabatic invariant, for different kinetic energy bins (see the labels). Results for the different populations are reported (from left to right): stably-trapped, quasi-trapped, un-trapped and the total under-cutoff proton sample. }
\label{Fig2}
\end{figure}
\begin{figure}[!t]
\centering
\includegraphics[width=6.5in]{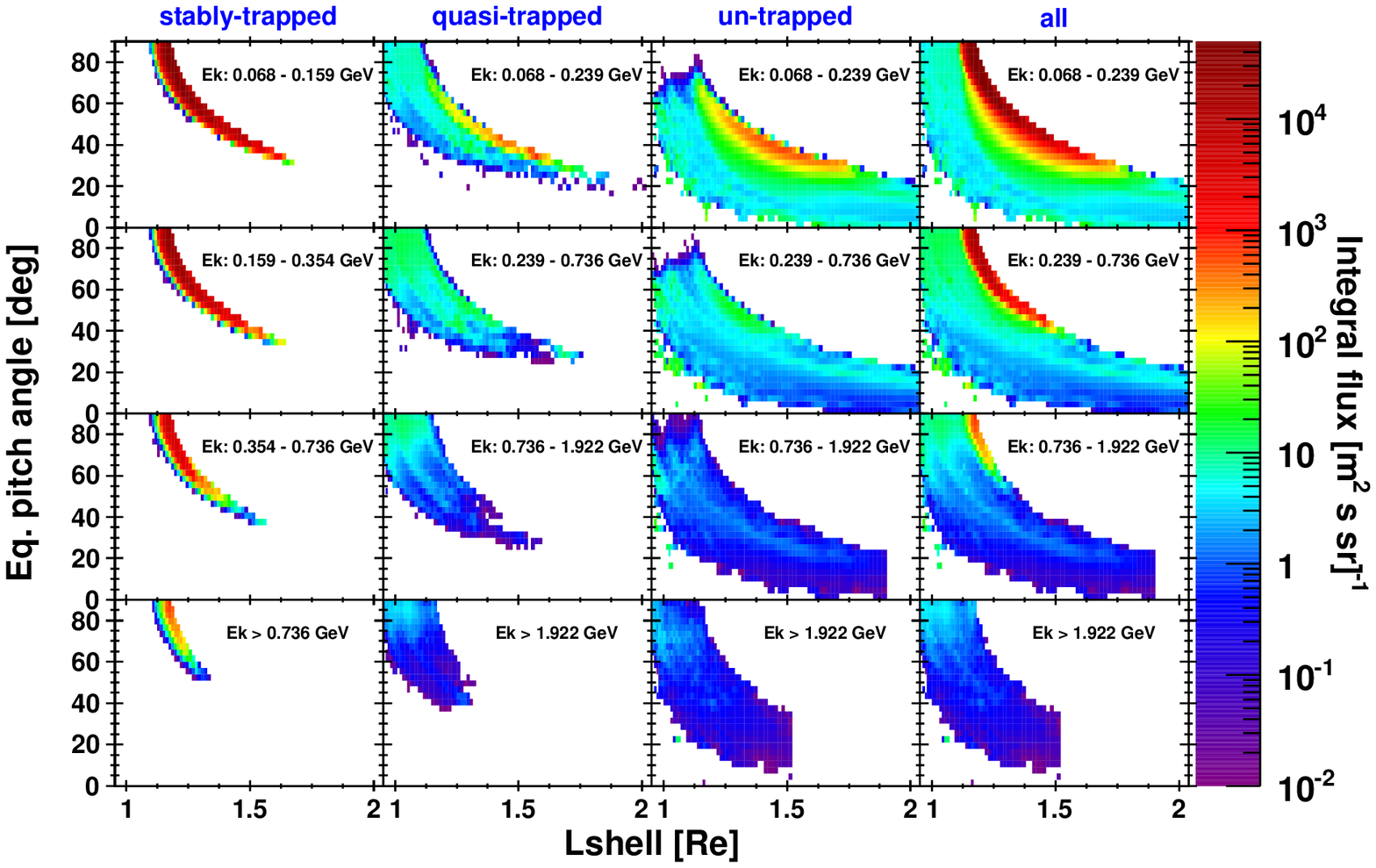}
\caption{The same as Figure \ref{Fig2}, but as a function of equatorial pitch angle $\alpha_{eq}$ and McIlwain's $L$-shell.}
\label{Fig3}
\end{figure}
Figure \ref{Fig2} shows the under-cutoff integral fluxes
as a function of adiabatic invariants $K$ and $\Phi$ for different kinetic energy bins. In order to improve resolution, $K^{1/2}$ and $log_{10}(\Phi)$ bins were used. The Y-axis ($K$ = 0) corresponds to the magnetic equator. Maps for the several populations (stably-, quasi- and un-trapped protons together with the entire under-cutoff proton sample) are reported.
Similarly, Figure \ref{Fig3} shows fluxes as a function of equatorial pitch angle $\alpha_{eq}$ and $L$-shell.

The first column in each figure reports results for stably-trapped protons.
Constrained by the spacecraft orbit, the covered phase-space regions varies with the magnetic latitude.
In particular, PAMELA can observe equatorial mirroring protons only for $L$-shell values up to $\sim$1.18 $R_{E}$ (or, equivalently, down to $\Phi$ $\sim$ 0.2 $G\cdot R_{E}^{2}$), and
measured distributions that result in strips of limited width parallel to the ``drift loss cone'' which delimits the $\alpha_{eq}$ (or $K$) range for which stable magnetic trapping does not occur.
Fluxes exhibit high angular and radial dependencies.

For a comparison, the second and the third columns in Figures \ref{Fig2} and \ref{Fig3} show fluxes for quasi- and un-trapped components.
Measured maps result from the superposition of
distributions corresponding to regions characterized by a different local (or bounce) loss cone value, i.e.,
the altitude of
the mirror points changes with the longitude drift,
decreasing from the SAA to the region on the opposite part of the planet
(sometimes called ``SouthEast-Asian Anomaly'' or SEAA), the closest to the eccentric dipole center, where the geomagnetic field has a local maximum.
Fluxes are quite isotropic, except for the SAA, where distributions are similar to those of stably-trapped protons;
conversely, energy spectra outside the SAA are harder and extend
to higher energies ($\lesssim$ 7 GeV),
especially in the SEAA, which is characterized by stronger trapping of energetic particles. Note that the un-trapped flux suppression at highest energy and $L$ (or, equivalently, lowest $\Phi$) bins is due to the cutoff selection used ($R<10/L^{3}$).

\subsection{Comparison with Models}

\begin{figure}[!t]
\centering
\includegraphics[width=6.5in]{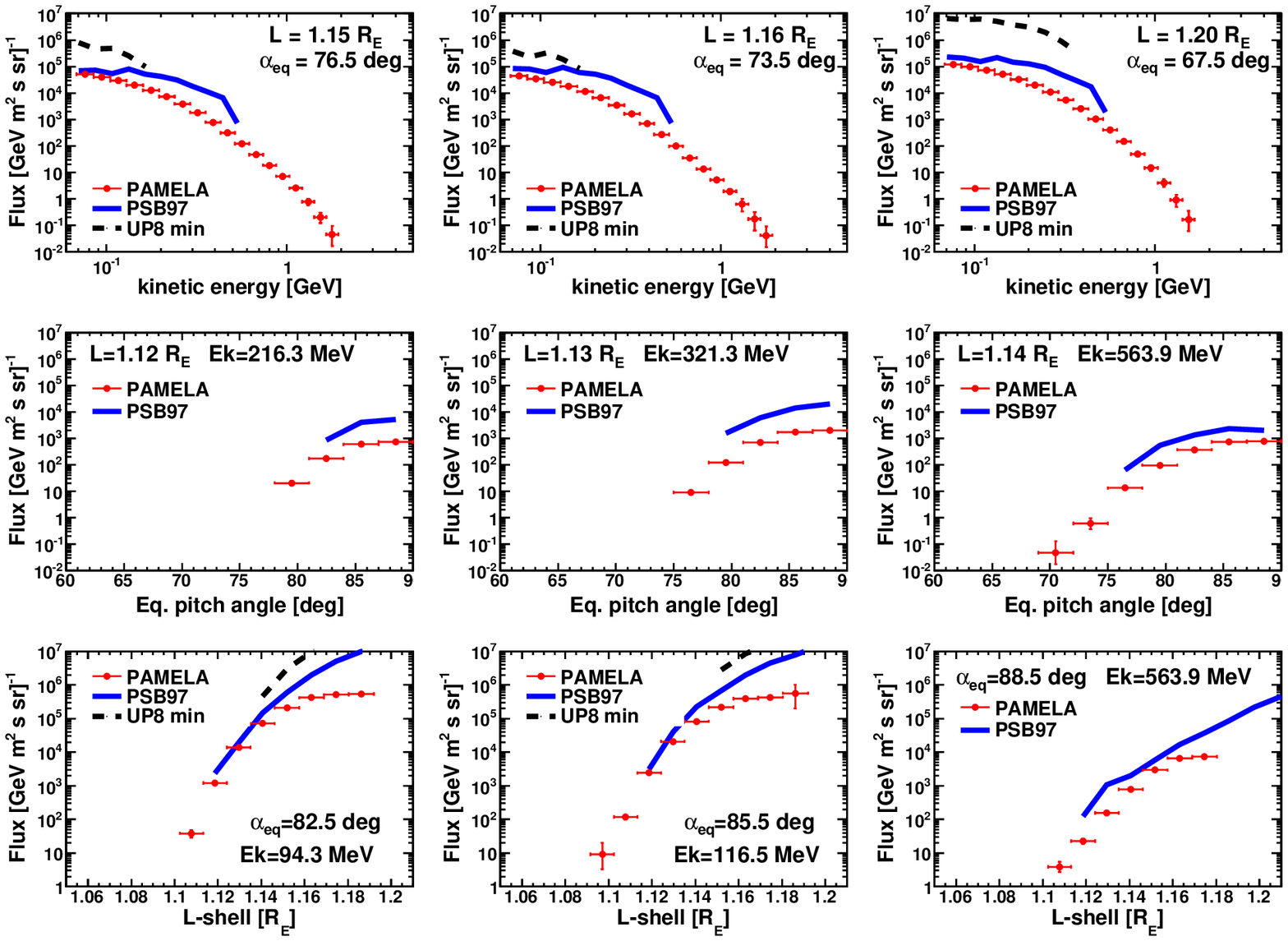}
\caption{Trapped proton energy spectra (top panels), pitch angle profiles (middle panels) and $L$-shell profiles (bottom panels) compared with predictions
from AP8-min model \citep{AP8} adapted for unidirectional fluxes (UP8-min) and from PSB97 \citep{PSB97}
model, denoted with dashed black and solid blue lines respectively.
Model data are from the SPENVIS web-tool \citep{SPENVIS}.
Comparisons are reported for combinations of sample $E_{k}$, $\alpha_{eq}$ and $L$-shell values.
}

\label{Fig4}
\end{figure}

Figure \ref{Fig4} compares PAMELA results and the predictions from two empirical models available in
the same energy and altitude ranges.
The former is the NASA AP8 model for solar minimum conditions, covering the energy range 0.1$\div$400 MeV \citep{AP8}; originally based on omnidirectional maps, it was adapted for unidirectional fluxes (UP8-min), and it implements the interpolation scheme by \citet{DALY}. The latter is the PSB97 model, based on data from the SAMPEX/PET mission: it provides directional fluxes covering the energy range 18$\div$500 MeV \citep{PSB97}.
Data were derived by using the
SPENVIS web-tool \citep{SPENVIS}.
Both models were constructed
following McIlwain's original procedure with a standardized dipole moment $M_{d}$ = 0.311653 $G \cdot R_{E}^{3}$ independent of epoch \citep{McIlwain}.

PAMELA results extend the observational range for trapped protons down to $L$ $\sim$ 1.1 $R_{E}$, and up to the maximum proton kinetic energies corresponding to trapping limits (critical energies $\lesssim$ 4 GeV).
Reported vertical bars account only for statistical errors; the total PAMELA systematic uncertainty amounts to 10$\div$20 \%, decreasing with increasing energy.
UP8 predictions deviate from PAMELA points almost everywhere, overestimating fluxes by about one order of magnitude.
Instead, a better agreement can be observed between the PAMELA and PSB97 predictions.
While pitch angle dependencies (Figure \ref{Fig4}, middle panels) appear to be consistent,
significant deviations from the models can been noted between energy spectra (Figure \ref{Fig4}, top panels); in fact, PAMELA fluxes
do not show the structures
present in the PSB97
predictions, resulting in a disagreement which amounts up to an order of magnitude at the highest energies.
Discrepancies in radial shapes at highest $L$-shell bins (Figure \ref{Fig4}, bottom panels) are due to
the limited
data
resolution at the highest spacecraft altitudes, which
affects the PAMELA flux
intensity estimate.

\begin{figure}[!t]
  \centering
  \includegraphics[width=4in]{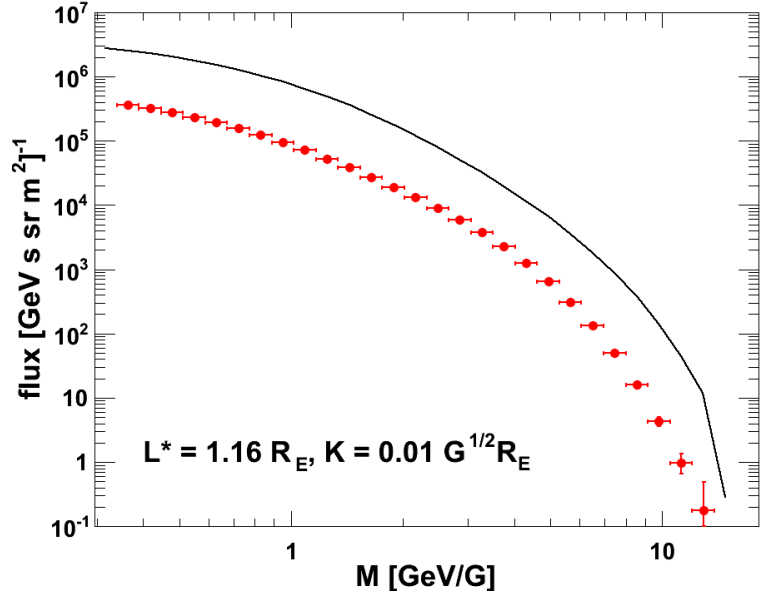}
  \caption{Stably-trapped differential flux ($GeV^{-1} m^{-2} s^{-1} sr^{-1}$) at geomagnetic equator compared with a theoretical calculation by \citet{Selesnick2007} for the year 2000. Spectra are reported as a function of the first adiabatic invariant $M$ for sample values of $K$ and $L^{*}$ invariants.}
  \label{Fig5}
 \end{figure}

Finally, Figure \ref{Fig5} shows the comparison between PAMELA
results and a theoretical calculation by \citet{Selesnick2007} for the year 2000. Differential fluxes are reported as a function of the first adiabatic invariant $M$, for sample values of $K$ and $L^{*}$ invariants (equatorial region). While spectral shapes are in a good qualitative agreement,
measured flux intensities end up being up to
about an order of magnitude
lower with respect to
model predictions,
depending on the phase-space region.

\section{Conclusions}\label{Conclusions}
The PAMELA measurements of energetic ($\gtrsim$ 70 MeV) geomagnetically trapped proton fluxes at low Earth orbits (350$\div$610 km) have been presented.
The analyzed sample corresponds to data acquired by PAMELA between 2006 July and 2009 September. Trajectories of selected events were reconstructed in the magnetosphere by means of a tracing code based on the numerical integration of particle motion equations in the geomagnetic field, and investigated in the framework of the adiabatic theory.

Stably-trapped protons
were detected in the South Atlantic Anomaly region, where the inner Van Allen belt makes its closest approach to the Earth's surface.
Measured spectra were compared with predictions of empirical and theoretical models available in
the same energy and altitude ranges.
The PAMELA results extend the observational range for the trapped radiation down to lower $L$-shells ($\sim$ 1.1 $R_{E}$) and up to the highest kinetic energies ($\lesssim$ 4 GeV), significantly improving the description of the low-altitude radiation environment
where current models suffer from the largest uncertainties.

\section*{Acknowledgements}
We acknowledge support from The Italian Space Agency (ASI), Deutsches Zentrum f\"ur Luftund Raumfahrt (DLR), The Swedish National Space Board, The Swedish Research Council, The Russian Space Agency (Roscosmos) and The Russian Scientific Foundation.
We gratefully thank R. Selesnick for helpful discussions, and D. Smart, M. A. Shea and J. F. Cooper for their assistance and support in adapting their trajectory program \citep{TJPROG}.

\end{document}